\documentclass{aa}
\input{psfig.sty}
\def\simgt{\lower.5ex\hbox{$\;\buildrel>\over\sim\;$}}
\def\simlt{\lower.5ex\hbox{$\;\buildrel<\over\sim\;$}}
\begin{document}


   \thesaurus{06     
              (11.01.2;  
               11.09.1;  
               11.09.2;  
               11.14.1;  
               11.19.1)} 
\authorrunning{Vanzi et al.}
\titlerunning{IRAS~19254-7245}

   \title{Multi-wavelength Study of IRAS~19254-7245 - The Superantennae \thanks{Based on observations obtained at the ESO telescopes of La Silla}}


   \author{L. Vanzi\inst{1}, S. Bagnulo\inst{1}, E. Le Floc'h\inst{1,2}, R. Maiolino\inst{3}, E. Pompei\inst{1}, W. Walsh\inst{4}}

   \offprints{lvanzi@eso.org}

   \institute{European Southern Observatory (ESO), 
              Alonso de Cordova 3107, Santiago - Chile
          \and
             present address CEA, Service d'Astrophysique 91191 Gif-sur-Yvette - France
          \and
              Osservatorio di Arcetri, Largo E. Fermi 5, 50125 Firenze - Italy
          \and
              Harvard Smithsonian Center for Astrophysics, 60 Garden St, MS 12, Cambridge, MA, 02138 - USA
             }

\date{Received ....; accepted ....}

\maketitle

\begin{abstract}

We present observations in the optical, near-infrared and millimetre
bands of the Ultraluminous Infrared galaxy IRAS~19254-7245, also known
as "The Superantennae".  This galaxy is an interacting system with a
double nucleus and long tails extending for about 350~kpc. We studied
in detail the southern component of the system which is optically
classified as a Sy2 galaxy. We have developed a method to determine
the parameters of the emission lines in a spectrum in the case of multiple
components and severe blending. Our data allow us to build a picture of
the environment around the nucleus of the galaxy with unprecedented
detail. The optical lines show a complex dynamical structure that is
not observable in the near-infrared.  In addition we find typical
features of AGN such as the coronal lines of [FeVII]5721 and
[SiVI]1.96. We also detect strong emission from [FeII]1.64 and $H_2$.

   \end{abstract}

%

\section{Introduction}

IRAS~19254-7245, with an infrared luminosity
$L_{IR}=1.1~10^{12}~L_{\odot}$ (calculated assuming $H_0=75 
Kms^{-1}Mpc^{-1}$, v=18479 $Kms^{-1}$ and using all four IRAS fluxes
Melnick \& Mirabel 1990), lies amongst
the Ultraluminous Infrared Galaxies (ULIRGs) emitting most of their
energy in the far infrared.  More than 90\% of these objects show
evidence for disturbed morphologies and interaction ( e.g. Sanders et al.
1988; Melnick \& Mirabel 1990; Murphy et
al. 1996; Clements et al. 1996). IRAS~19254-7245 is a particularly spectacular
interacting galaxy with tidal tails extending to a distance of about
350~kpc and two bright nuclei 10~kpc apart.  Although it is widely
accepted that this system results from the collision between two
gas-rich spirals, a multiple-merger origin has been recently suggested
from high-resolution HST observations showing that a double nucleus
may be present in both components of the interaction (Borne et
al. 1999).

The southern nucleus prevails in luminosity as the system is observed
at longer wavelengths and it is the dominant source in the
mid-infrared as shown by the ISOCAM observations of Charmandaris et
al. (2002).  This nucleus also shows an optical spectrum typical of a
Seyfert 2 galaxy (Mirabel et al. 1991).  The presence of strong
nuclear activity is moreover indicated by the Near-Infrared (NIR) and IRAS
25/60\footnote{defined as Log f$_{25}$/f$_{60}$} colors, which both
suggest the presence of important thermal emission from hot dust
($\sim$ 300-500\,K) most probably heated by the active nucleus.

The system has been studied among others by Mirabel et al. (1991) and
Colina et al. (1991).  Both groups find broad emission lines in the
optical with a complex profile that can be attributed to
material falling onto the nucleus or to an outflow.
The kinetic energy required to power such a
flow may originate from stellar winds and/or
supernovae. This strongly supports the picture of merger-driven
starburst activity characterized by a high star formation rate ($\sim$
150 M$_{\odot}$\,yr$^{-1}$, Colina et al. 1991), taking place in the
Seyfert circumnuclear region.

Understanding the true origin of the ultra-luminous phase in infrared
galaxies is still an open debate.  Recent studies have shown that the
merger-stage in interacting systems, traced by the projected
separation of the nuclei, correlates with the star-forming efficiency
and with the molecular mass detected in each merger (Gao \& Solomon
1999; Murphy et al. 2001). However the relative importance between
starbursting and AGN activity in luminous galaxies, and ULIRGs in
particular, is still an open issue (Sanders \& Mirabel 1996,
Lutz et al. 1996, Vignati et al. 1999).
Harboring both
a Seyfert and a starburst components, IRAS 19254-7245 is a very
interesting object within this context. The separation between the two
nuclei involved in this merger is still large enough to allow a
careful analysis and
distinguish between the two sources. Our main goal is to study the near
infrared (NIR) spectrum of this galaxy, but optical and $mm$ data were
also obtained.
 
The present paper is divided in three main sections: Observations,
Analysis and Discussion. In $\S$~2 we present the data describing how
they were obtained and reduced. In $\S$~3 we present the analysis of
the data and how the main physical parameters were derived from the
observations. In $\S$~4 we use these parameters to describe the
physical conditions in the galaxy. Finally we summarize our
conclusions.


\section{Observations}

\subsection{Optical}

\begin{figure}
\psfig{figure=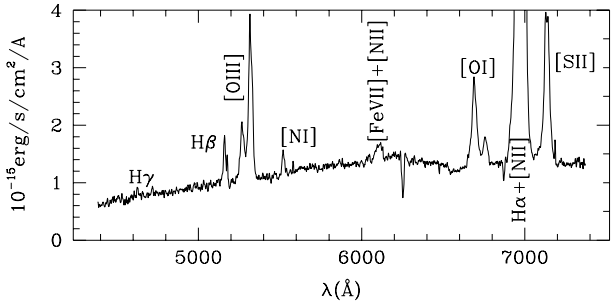,width=8.5cm,angle=0}
\caption{Low resolution optical spectrum of IRAS~19254-7245 obtained at the 
1.5 ESO telescope of La Silla.}
\label{eso}
\end{figure}

We obtained low and medium resolution spectra of IRAS~19254-7245 at
optical wavelengths. The low resolution spectrum was observed with the Boller \& Chivens
spectrograph at the ESO 1.52m telescope at La Silla in 1999 May using
grating \#23 and a 2\arcsec\,-wide slit to yield a spectral resolution of
R=1300 in the range 4400-7400\AA. The position angle was --12\degr, which
allows to include both nuclei
in the slit. The total integration time
was 40 min. The 1D spectrum of the southern component extracted with an
aperture of 2.6\arcsec\, corresponding to 3.1 kpc at a
distance of 247 Mpc, is shown in Fig.\ref{eso}.  The medium resolution 
spectrum was acquired with EMMI at the ESO-NTT in 1999 July using the
REMD (Red Medium Dispersion) mode and grating $\#$6 which gives a
resolution R=5500 in the band 6600-7200\AA. Two exposures of 15 and
30 minutes were taken using a 1\arcsec\,-wide slit.  HeAr lamp spectra were
taken before and after the exposures for the wavelength
calibration and the
star EG 274 was observed as spectro-photometric standard.  The spectra
were reduced following standard procedures.  The EMMI instrumental
resolution, estimated on sky lines is 36~km~s$^{-1}$.  The flux
calibration was obtained by comparing the wavelength calibrated
spectra of the standard star with the flux table published.  
From the 2D combined frame we extracted 1D spectra with an aperture of 2.6\arcsec. 
A linear fit on the continuum gives a slight positive slope from $9.4~10^{-17}$ at 
6600\AA to $9.9~10^{-17}~\mathrm{s}^{-1}\mathrm{cm}^{-2}$\AA$^{-1}$ at 7200\AA. 
To double check the flux calibration we used a
HST archive image taken in the F814W filter centered at 7940\AA. For
our spectroscopic aperture we obtained a flux of
$1.09~10^{-16}~\mathrm{erg}\mathrm{s}^{-1}\mathrm{cm}^{-2}$\AA$^{-1}$,
that agrees with our calibration to better than 5\%.
The medium resolution spectrum 
of the southern nucleus subtracted of the continuum is shown in
the top panel of Figs.\ref{emmi_free} and \ref{emmi_fix}.

\subsection{Near-Infrared}

We observed two spectra with the medium resolution grism of SOFI at
the ESO-NTT.  The spectra are respectively centered on the H and the
Ks bands.  They were acquired in 1999 May with 40 minutes of integration in
H and 66 minutes in Ks. The Ks spectrum was observed again in 1999
September for 30 minutes of integration.  We always used a 1\arcsec\,-wide slit
giving a resolution R=900 in H and R=1350 in Ks.  A low resolution
spectrum (R=600) covering the range from 1.5 to 2.5 $\mu m$ has been
obtained with SOFI in 2001 June, the total integration time was in
this case 40 minutes.  The position angle was always --12\degr.  The data
reduction followed the standard steps for NIR spectroscopy.
1D spectra were extracted with an aperture of 2.6\arcsec\, for the
southern component. Atmospheric
features were corrected by dividing for the spectrum of a reference
star. In the observation of 1999 May we used a G1V and a F8V star,
then the spectra were multiplied by the solar spectrum to remove the
stellar features and reestablish the correct slope of the
continuum (Maiolino et al. 1996). 
In 1999 September we used an O6 star and then multiplied by
a blackbody at 40000 K to reestablish the correct slope of the
continuum.
The Ks spectra observed on the two occasions give a good
check on the reliability of this method; they are virtually identical
showing no residual features from the reference star and exactly the
same slope of the continuum. The Ks spectra have been averaged
together giving a final spectrum with 1h36m of integration.  To flux
calibrate the spectra we obtained two images in the H and Ks band with
SOFI in 2001 June; the integration time was 15 minutes in each band. From
the images we extracted the photometry on the spectroscopic aperture
obtaining fluxes of $3.16~10^{-12}$ and $2.80~10^{-12}~$erg~s$^{-1}$~cm$^{-2}\mu$m$^{-1}$
in H and Ks respectively. The photometric error was always below 3\%.
This value does not include the uncertainty on the spectroscopic
apertures and on their centering, however we believe that
this procedure guarantees a very accurate flux
calibration. In the case of the low resolution spectrum the slope of
the continuum was readjusted to match the photometric points. In
Fig.\ref{sofi} we present the NIR spectra of the southern nucleus. The
top curve is the low resolution spectrum calibrated in flux. In the
middle the medium resolution spectrum.  In the bottom part of the left
panel we show a stellar template redshifted to match the galaxy with
the indication of the absorption features detectable. In the
bottom right panel the K spectrum, with the $\mathrm{H}_2$(1-0)S(3)
line subtracted, clearly shows the detection of [SiVI] (see Section 3.2
for details).  Our NIR spectra of the northern nucleus are virtually
featureless.

\subsection{Millimetre}

The radio observations were performed with the Swedish
ESO-Submillimetre Telescope (SEST) in La Silla
during 1999 November.  The FWHM beam sizes are 45\arcsec\, and
23\arcsec\, at 115GHz and 230GHz respectively, and the main beam
efficiencies at these frequencies are 0.70 and 0.50. Intensity
calibration is done with the chopper wheel method, so the raw data, in
units of T$^{\rm{A}}_*$, are divided by the main beam efficiency to
obtain main beam brightness temperatures. The internal consistency of
the SEST is accurate to within a few percent.

The backends used were two acousto-optical spectrometers, each with a
total bandwidth of 1 GHz. System temperatures ranged between 200K and
350K on the T$^{\rm{A}}_*$ scale. All observations of the CO lines were
performed with $\tau_{\rm 225GHz} \leq 0.4$. The SEST's absolute
pointing accuracy is 3\arcsec\ rms in azimuth and elevation, and the
pointing model was checked with regular observations of SiO maser
sources.
Beam switching mode was employed, where the secondary mirror was
wobbled with a beam throw of 697\arcsec\ in azimuth and scans obtained with
reference positions on either side were coadded to ensure flat
baselines. The data were reduced with the CLASS software of the
Grenoble Astrophysical Group (GAG) package. A polynomial baseline of
order one, or occasionally three, was removed from each spectrum
before averaging. The spectra are shown in Fig. 4.

\begin{figure}
\psfig{figure=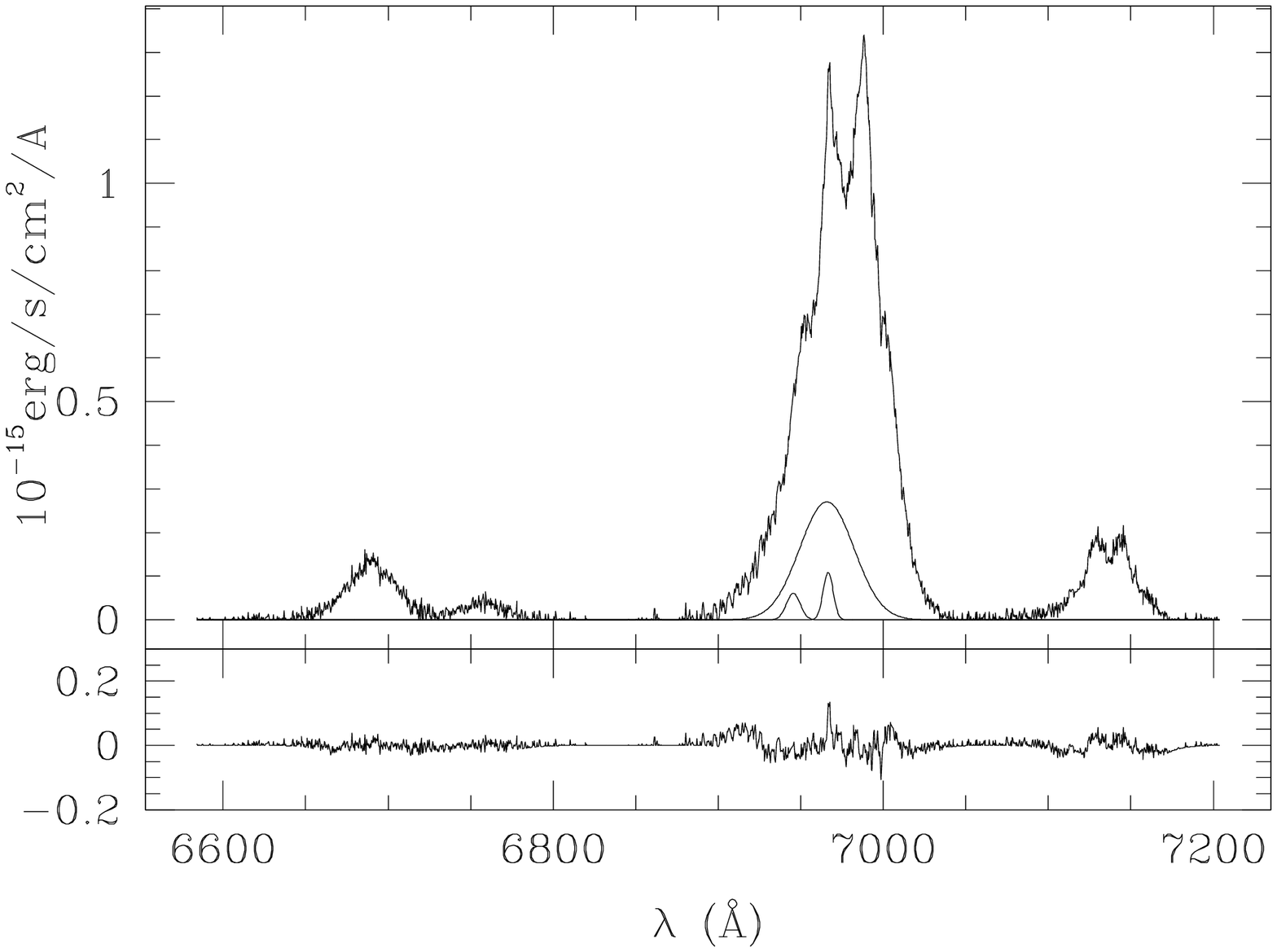,width=8.2cm,angle=0}
\caption{Medium resolution optical spectrum of IRAS~19254-7245 observed with EMMI at the
ESO-NTT. Top panel: the spectrum subtracted of the continuum and the three components of
H$\alpha$ obtained with free $\sigma$. Bottom panel: the residuals of the fit.}
\label{emmi_free}
\end{figure}

\begin{figure}
\psfig{figure=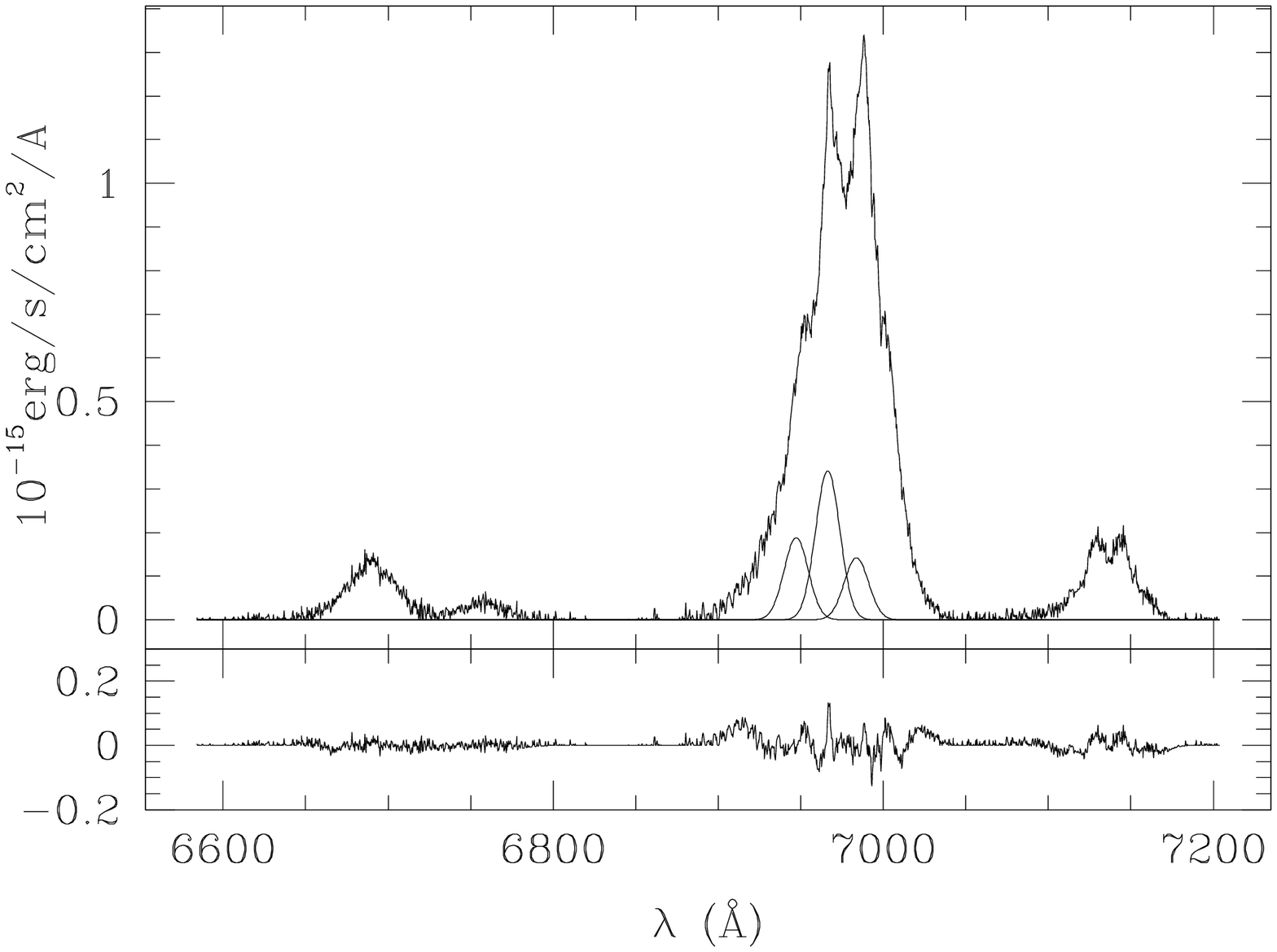,width=8.2cm,angle=0}
\caption{Medium resolution optical spectrum of IRAS~19254-7245 observed with EMMI at the
ESO-NTT. Top panel: the spectrum subtracted of the continuum and the three components of
H$\alpha$ obtained with fixed $\sigma$. Bottom panel: the residuals of the fit.}
\label{emmi_fix}
\end{figure}

\begin{figure*}
\psfig{figure=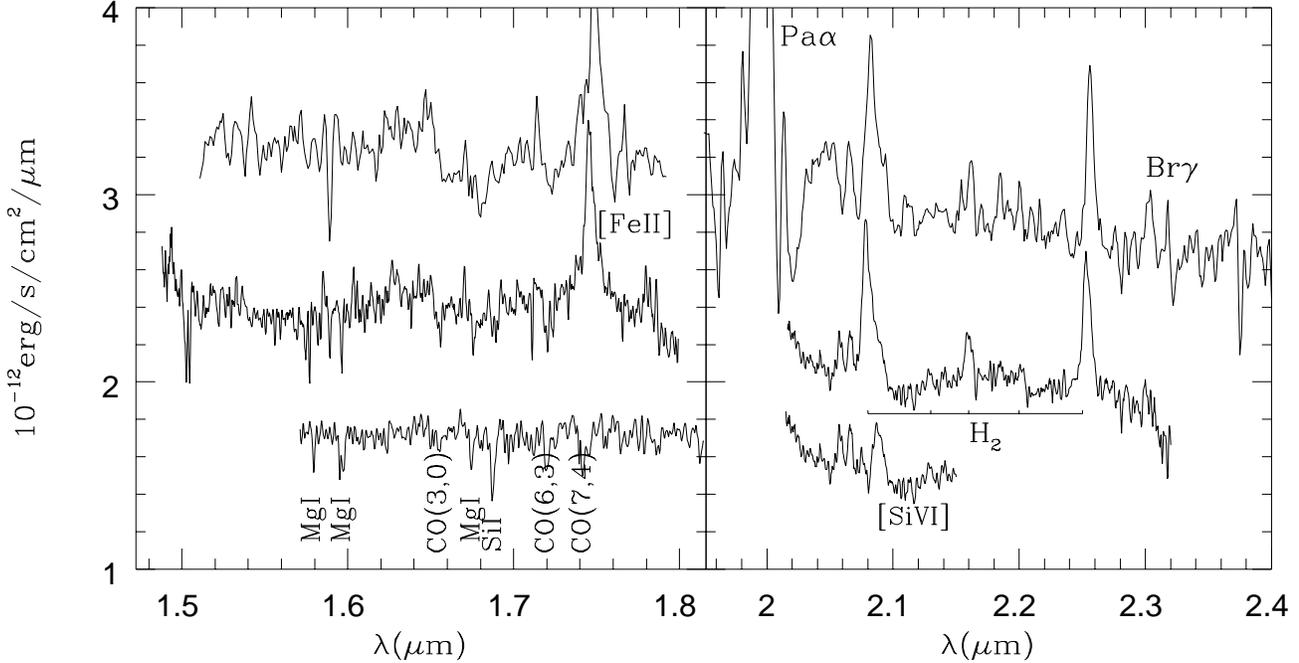,width=18cm,angle=0}
\caption{NIR spectrum of IRAS~19254-7245 observed with SOFI at the ESO-NTT. From top low 
resolution, medium resolution, stellar template on the left and spectrum subtracted of the
(1-0)S(3) line on the right. The flux scale only refeers to the low resolution spectrum.}
\label{sofi}
\end{figure*}

\section{Analysis}

\subsection{The Optical Spectrum}

The optical spectrum of the southern nucleus shows broad emission
lines even at low resolution. At high resolution it is evident how
each emission line contains more than one component.
It is not an easy task to describe the lines in terms of simple components,
since most of the
lines in the observed
spectral range observed are blended.
After several unsatisfactory attempts with the most common data
reduction packages we decided to develop our own code to produce the
best fit of the observed spectrum taking into account all known
constraints, mainly [OI]6363 = 0.33 [OI]6300 and [NII]6583 = 3 [NII]
6584.

\begin{table*}
\caption{Best-fit parameters for the 1, 2, and 3 component models. Fluxes are in units of $10^{-15}$ \,erg\,cm$^{-2}$\,s$^{-1}$. See text for details. }
\begin{tabular}{lccccc|cccc}
\hline
 & &
\multicolumn{4}{c|}{Low Resolution}&
\multicolumn{4}{c}{High Resolution}\\
\multicolumn{2}{l}{Parameter}
             & 1-Comp.&  2-Comp.&\multicolumn{2}{c}{3-Comp.}&
                                                 1-Comp.&  2-Comp.&\multicolumn{2}{c}{3-Comp.}\\
   &         &  Model  & Model&\multicolumn{2}{c}{Model}    &   Model & Model&\multicolumn{2}{c}{Model}\\
             &         &        &         &         &         &         &         &         &           \\
\hline
$F_1$        &         & 100\,\%&  88\,\% &  82\,\% &55.0\,\%& 100\,\%&   91\,\%&  88.5\,\% &51.0\,\%   \\
$F_2$        &         &        &  12\,\% &  12\,\% &22.0\,\%&        &    9\,\%&   5.0\,\% &28.0\,\%   \\
$F_3$        &         &        &         &   6\,\% &23.0\,\%&        &         &   6.5\,\% &21.0\,\%   \\
\hline
$z_1$        &         & 1.06153& 1.06165 & 1.06196 &1.06148&  1.06132&  1.06144&  1.06141&1.06149  \\
$z_2$        &         &        & 1.06148 & 1.06146 &1.05850&         &  1.06164&  1.05831&1.05858  \\
$z_3$        &         &        &         & 1.05422 &1.06415&         &         &  1.06153&1.06414  \\
\hline
$\sigma_{H\alpha\,1}$& &21.1\,\AA&24.4\,\AA&22.0\,\AA&10.4\,\AA&22.2\,\AA&23.7\,\AA&23.3\,\AA&10.4\,\AA\\
$\sigma_{H\alpha\,2}$& &         & 5.7\,\AA& 5.6\,\AA&10.4\,\AA&         & 4.7\,\AA& 5.8\,\AA&10.4\,\AA\\
$\sigma_{H\alpha\,3}$& &         &         &12.6\,\AA&10.4\,\AA&         &         & 4.3\,\AA&10.4\,\AA\\
\hline
H$\beta$ &(4861\,\AA)  &  1.46  &  1.52   &   1.69  & 1.61  &         &         &         &           \\
O\,III   &(4959\,\AA)  &  2.36  &  2.36   &   2.28  & 2.49  &         &         &         &           \\
O\,III   &(5007\,\AA)  &  7.10  &  7.24   &   7.58  & 7.19  &         &         &         &           \\
N\,I     &(5198\,\AA)  &  0.61  &  0.63   &   0.80  & 0.75  &         &         &         &           \\
Fe\,VII  &(5721\,\AA)  &  0.26  &  0.27   &   0.10  & 0.10  &         &         &         &           \\
N\,II    &(5755\,\AA)  &  1.06  &  1.05   &   0.87  & 0.87  &         &         &         &           \\
O\,I     &(6300.3\,\AA)&  3.59  &  3.77   &   4.63  & 4.09  &   3.70  &   3.89  &  3.43   & 3.44    \\
S\,III   &(6312.1\,\AA)&  0.89  &  0.78   &   0.77  & 1.24  &   1.37  &   1.15  &  1.57   & 1.42    \\
O\,I     &(6363.8\,\AA)&  1.18  &  1.24   &   1.53  & 1.35  &   1.22  &   1.28  &  1.14   & 1.14    \\
Fe\,X    &(6375.0\,\AA)&        &         &   0.25  & 0.38  &         &         &  0.27   & 0.24    \\
N\,II    &(6548.1\,\AA)& 10.94  & 10.49   &  10.03  &11.01  &  12.8   &  12.1   & 13.47   &13.26     \\
H$\alpha$&(6562.8\,\AA)& 19.35  & 22.0    &  25.28  &18.75  &  13.7   &  17.1   & 12.59   &12.33    \\
N\,II    &(6583.4\,\AA)& 32.82  & 31.5    &  30.09  &33.03  &  38.4   &  36.3   & 40.42   &39.77    \\
S\,II    &(6716.4\,\AA)&  7.01  &  7.15   &   7.76  & 6.57  &   3.38  &   4.05  &  3.12   & 3.16    \\
S\,II    &(6730.8\,\AA)&  3.39  &  3.43   &   2.79  & 3.58  &   5.16  &   4.56  &  4.10   & 3.87    \\
\hline\\
\end{tabular}
\label{bagu_tab}
\end{table*}
 
We considered three different cases: a 1-component model, a
2-component model and a 3-component model. For the 3-component model
we have considered 2 cases, one with free $\sigma$ for the lines, the other
with fixed $\sigma$. The latter gives a more physical result given the type 2
nature of the Seyfet galaxy. We applied the inversion
code to both the low resolution and the high resolution spectra.  The
best-fit parameters are given in Table~1, which is organized as
follows. $F_i$ denotes the fraction of the flux ascribed to the $i$-th
component of the line, $z_i$ denotes the red-shift, $\sigma_{H\alpha\,i}$ the
standard deviation of the Gaussian of the $i$-th component 
and refers to the H$\alpha$ line. It
should be noted that the contribution due to the instrumental
resolution, $\sigma_\mathrm{instr} \simeq 5$\AA\ (FWHM), and 1.1\AA\ (FWHM), for the
low and the high resolution spectrum, respectively, is negligible
compared to the intrinsic width of each component.
The remaining rows in the table
give the line integrated flux, due to \textit{all} components of the
various emission lines, in units of
$10^{-15}$\,erg\,cm$^{-2}$\,s$^{-1}$.  The residuals of the best fit
to the high-resolution spectrum obtained assuming a 
3-component model with free and fixed $\sigma$ are shown in
Figs.\ref{emmi_free}
and \ref{emmi_fix}, along with the components of $H\alpha$.

The single component model is clearly insufficient to explain the
observations.  A much more satisfactory fit was obtained by assuming a
2-component model. The best results were obtained by assuming a
3-component model, and even better results would be expected by adding
a fourth component (as done by Colina et al.\ 1991).  However,
increasing the number of free parameters leads to multiple solutions
with similar values of the minimum $\chi^2$, which makes it virtually
impossible to determine the most reliable one. 
The scattering between the various values obtained from the different
models allows us to give the most appropriate estimate of the
errors associated with the parameters, i.e., typically 3\% for
the high resolution spectrum.

\begin{figure}
\psfig{figure=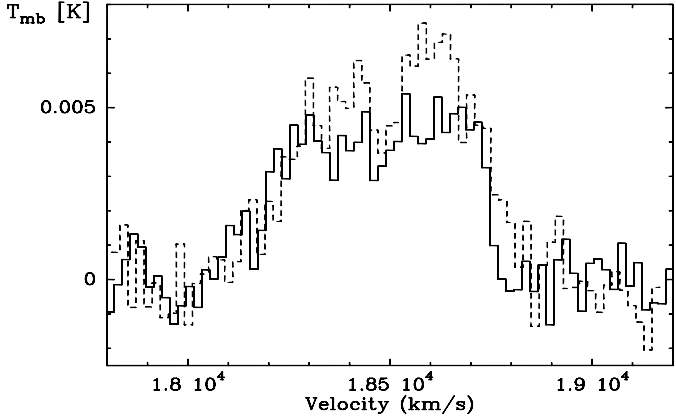,width=9cm,angle=-90}
\caption[]{SEST $^{12}$CO(1-0) (solid line) and $^{12}$CO(2-1) (dotted line)
spectra of IRAS~19254-7245. The data have been smoothed to a velocity
resolution of 20~km~s$^{-1}$.}
\label{co}
\end{figure}

\begin{table}
\caption{NIR emission lines fluxes measured in 10$^{-15}$\,erg/s/cm$^2$. (*) have been
measured in the low resolution spectrum only. The error on P$\alpha$ is undetermined
due to the bad atmospheric transmission.}
\label{nir_lines}
\begin{tabular}{ccccc}
\hline
line     & $\lambda_{obs}$ & $\lambda_{rest}$ & Flux & FWHM (km~s$^{-1}$) \\
\hline
[FeII]         & 1.746 & 1.644 & 5.9$\pm$ 0.1 &  1200 \\
P$\alpha$(*)  & 1.993 & 1.875 & 38$\pm$ ? &   630 \\
Br~8           & 2.067 & 1.945 & 1.0$\pm$ 0.3 &   400 \\
$[SiVI]$       & 2.087 & 1.962 & 2.2$\pm$ 0.3 &  1180 \\
Br$\gamma$(*)  & 2.300 & 2.165 & 2.0$\pm$ 0.5 &    -  \\
\hline
$\mathrm{H}_2$ line     &       &       &           &       \\
\hline
(1-0)S(3) & 2.079 & 1.957 & 5.8$\pm$ 0.2 &  1070 \\
(2-1)S(4) & 2.129 & 2.003 & 0.4$\pm$ 0.2 &    -  \\
(1-0)S(2) & 2.160 & 2.034 & 1.6$\pm$ 0.2 &   765 \\
(2-1)S(3) & 2.200 & 2.070 & 0.3$\pm$ 0.2 &    -  \\
(1-0)S(1) & 2.254 & 2.121 & 5.2$\pm$ 0.2 &   880 \\
\hline
\end{tabular}
\end{table}

\subsection{The NIR Spectrum}

The NIR spectrum has a resolution that is significantly lower than the
optical medium resolution spectrum and not enough to discriminate the 
various velocity components. For this reason it has not been necessary to apply
our code to the NIR part of the spectrum and all lines have been fitted
with a single gaussian component.

The [SiVI] line blended with the line (1-0)S(2) of $\mathrm{H}_2$ is
clearly detected.  To make this detection more evident and to measure
the flux emitted we have subtracted from the (1-0)S(2) the profile
derived from the (1-0)S(1) line. The best subtraction is obtained
assuming a line ratio S(2)/S(1)=1.17.  The subtracted spectrum is
shown in the bottom part of Fig.\ref{sofi}; the two faint lines on the
left of [SiVI] can be identified with the residual of the OH 8-6
P1(4.5) sky line (Rousselot et al. 2000) and with Br$\delta$. In Table
\ref{nir_lines} the lines detected in the NIR spectrum and their
fluxes are listed.

\subsection{CO Spectrum}

Fig.\ref{co} shows the SEST spectra of the $^{12}$CO(1-0)
and $^{12}$CO(2-1) emission in IRAS~19254-7245. The spectra have
somewhat higher signal to noise than the $^{12}$CO(1-0) spectrum of
Mirabel et al. (1991), as our SEST spectra have rms $\sim 1.0$~mK when
averaged to $20$~km~s$^{-1}$ channels. Integrating these spectra we
find a total $^{12}$CO(1-0) flux of I(CO) = $2.4 \pm
0.1$~K~km~s$^{-1}$ and for $^{12}$CO(2-1) I(CO) = $3.0 \pm
0.1$~K~km~s$^{-1}$, where the uncertainties are estimated following
Elfhag et al. (1996). At the $\Delta v_{20}$ level (width of the line
at 20\% of its peak value), the line widths are $670\pm20$~km~s$^{-1}$
although there is a suggestion that the $^{12}$CO(2-1) spectrum is as
wide as the $800$~km~s$^{-1}$ claimed by Mirabel et al. (1991). The
width at 50\% level is $570~km~s^{-1}$. Both
lines have a central velocity consistent with the heliocentric radial
velocity of $18500 \pm 80$~km~s$^{-1}$ generally adopted for
IRAS~19254-7245. The differing beam sizes make line ratio
considerations impractical with the current data.

\section{Discussion}

\subsection{Extinction}

From the ratio of the hydrogen recombination lines it is possible to
calculate the extinction of the galaxy. Colina et al. (1991) derive
E(B-V)=1.40 from the ratio of $\mathrm{H}\alpha$ and $\mathrm{H}\beta$
that corresponds to $A_V=4.37$ assuming the extinction curve of Rieke
\& Lebofsky (1985). 
Using 3 components to deconvolve the $\mathrm{H}\alpha$ at high resolution
(see right column in Table~3) and the $\mathrm{H}\beta$ flux as measured at low resolution
we obtain $A_V=3.15$. We ascribe the difference with the value of Colina 
et al. to our better deconvolution of the spectrum.
For the theoretical ratio of the two lines we have used the value
2.86 given by Osterbrook (1989) for $T=10^4~K$ and $n_e=10^2~cm^{-3}$. 
An independent measure can be derived from the ratio
$\mathrm{H}\beta/\mathrm{Br}\gamma$. Using the theoretical ratios 36.9
we obtain $A_V=3.9\pm0.2$. This value, though less affected by the blending of the lines, 
has a much larger error. We do not use the $\mathrm{Pa}\alpha$ 
line that, though much
brighter than $\mathrm{Br}\gamma$, lies in a region of poor
atmospheric transmission and cannot be considered reliable as evident
from the high underlying noise visible in Fig.\ref{sofi}.

\subsection{Continuum}

The four photometric points available for our spectroscopic aperture,
mainly V, I, H and Ks, give a
rising continuum from the optical to the H band that significantly
flattens toward the Ks band.  We tried to reproduce the global shape
of the continuum using
a simple model including cold stars, warm dust and a non-thermal
continuum. The wide spectral range considered allows to put tight constraints
despite the relatively small number of measured points.
For our simplified model, each of these components was
described by a black body at 3000 K, a black body at 1500 K and a
power low with spectral index -0.5.  A good fit was obtained with a
56\% contribution from cold stars obscured by 4 mag of visual extinction,
28\% from dust with 4 mag of extinction and 16\% from the power-law.
The latter, nonthermal component is required to flatten the NIR part of
the spectrum; the same effect can however be obtained using 
a few hot stars heavily obscured or a cold star component with no 
extinction.

The NIR continuum of IRAS~19254-7245 appears rich of stellar
absorption features most of which are CO bands. We compared our
H-band spectrum with the spectrum of a stellar template obtained
combining the spectra of late type stars (see Engelbracht et
al. 1998), the result is shown in the bottom part of
Fig.\ref{sofi}. We run a $\chi^2$ test adding to the stellar template
an increasing percentage of flat featureless continuum to measure the
degree of dilution.  We found a minimum for the $\chi^2$ with a
dilution by about 30\%. Though this estimate is quite uncertain due to
the moderate S/N and resolution, it is certainly indicative of the
presence of some dilution of the continuum in the H band and it is not
too far from the previous estimate.
The absorption bands detected in the K band are too shallow to extend this
kind of analysis to this band.

The discrepancy in the contribution of the cold star component derived
with the two methods can be attributed to our over simplified model of
the continuum and to the poor detection of the stellar absorption
bands. We conclude that a significant fraction of
the continuum originates from a hot dust component, probably close
to the sublimation temperature, in
the vicinity or within the torus of the AGN. This is consistent with the
fact that a high fraction of the bolometric luminosity of the
SuperAntennae orginates from the nucleus of the southern galaxy as seen
in the Mid-Infrared (Charmandaris et al. 2002).

\subsection{Line Profiles}

As already noted, the optical emission lines show a complex structure
that cannot be reproduced by a simple gaussian fit. From the
analysis described in the previous section we have been able to obtain
a satisfactory fit describing each line as the superposition of two or
three components.  The observed profiles can be explained by the
presence of material that falls on or is ejected from the nucleus,
with typical velocities of about 1000~km~s$^{-1}$, as shown by the
parameters of the fit shown in Table \ref{bagu_tab}.

\begin{figure}
\psfig{figure=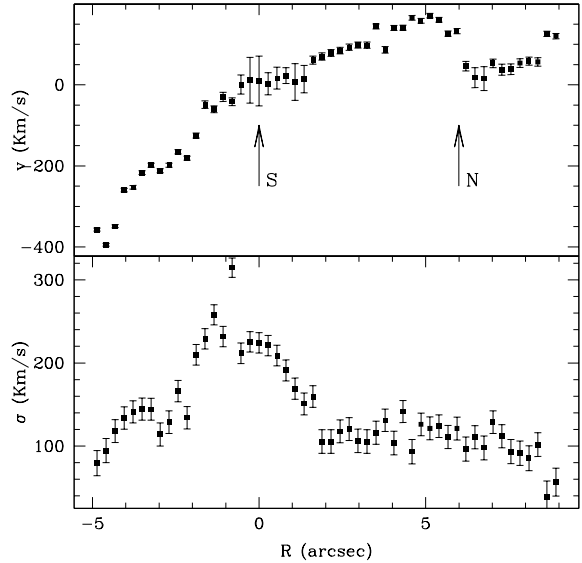,width=8cm,angle=0}
\caption{Rotation curve (upper panel) and velocity dispersion (lower panel)
of IRAS~19254-7245. The velocity is measured with respect to the southern component, the positions of the two nuclei are marked}
\label{velocity}
\end{figure}

On a larger scale, we have measured the rotation curve of the galaxy
from the $\mathrm{H}\alpha$ 2D spectrum. The curve is plotted in
Fig.\ref{velocity} with the southern nucleus at position 0. In this
scale the northern component appears to be located at about 6
arcsec. The positions of both nuclei are marked by arrows.
In the same figure we also plot the values of the velocity
dispersion ($\sigma$) deconvolved for the instrumental profile. The
rotation curve is consistent with that measured by Mirabel et al
(1991); however our spatial resolution is higher. At position between
8 and 9 arcsec we detect a cloud of gas with a velocity of at least
50 Km/s out of the main stream and a velocity dispersion lower than
the average.  This finding indicates that a dynamically complex
structure is present even at large radial distances.  The higher
errors in the velocity and the higher values of dispersion around
position 0 are mainly due to the complexity of the emission lines in
this region. All lines observed in the optical are "narrow" in
agreement with the classification of the south component as type 2.

It is interesting to note that the NIR lines do not show the complex
profiles detected in the optical and that they can be well fitted by single
gaussians. This is possibly due to the lower resolution and
lower S/N in the NIR that do
not allow a good detection of the faint components.
The velocity dispersion of the NIR lines are comparable to those in
the optical indicating that the optical and NIR lines are probably
generated in clouds that participate in the same motions. It would be
interesting to compare the redshift of the NIR lines with the optical
de-blended lines but this would require a spectral resolution and
wavelength accuracy higher than we have at the moment.

\subsection{Coronal Lines}

Two high ionization lines are detected in our spectra
[FeVII]5721 and [SiVI]1.962.  Both are blended with other lines
and a careful de-blending process was used to measure a reliable value
of the flux emitted. Marconi et al. (1994) proved that the ratio
[FeVII]5721/[SiVI]1.962 is very close to unity in case of
photoionization, independently of the details of the models. Correcting
the fluxes measured in these lines for $A_V=4$ with the extinction law
of Rieke \& Lebofsky (1985) we obtain [SiVI]/[FeVII] = 0.5. 
This value however is quite uncertain due to the large uncertainty on
the flux of the blended [FeVII]. In addition
the low resolution optical spectrum was
observed with a larger slit than the NIR one, 2\arcsec\, instead of
1\arcsec, and the [FeVII] could be diluted by the continuum.
Althought we have included the [FeX] line in our fit of the optical
spectrum there is no direct evidence for the detection of this line.

\subsection{Lines of $\mathrm{H}_2$ and [FeII]}

\begin{figure}
\psfig{figure=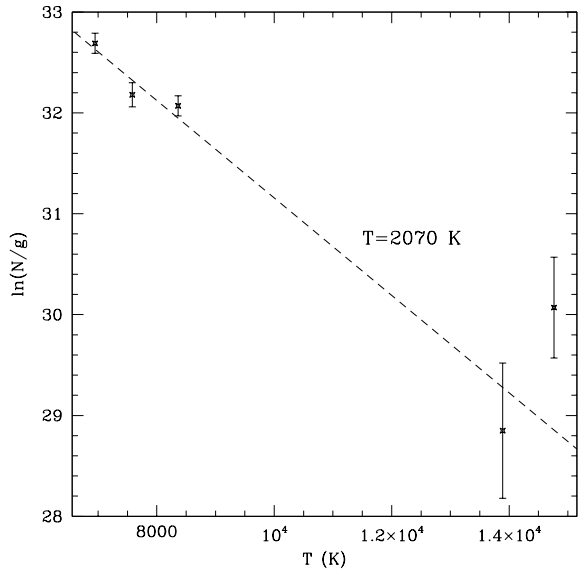,width=8cm,angle=0}
\caption{Excitation diagram of Molecular Hydrogen in IRAS~19254-7245.}
\label{exh2}
\end{figure}

We observe a very good NIR emission line spectrum from molecular
hydrogen. In Table 3 we compare the observed ratios with the values
predicted by two different models for fluorescent and thermal excitation
derived from Engelbracht et al. (1998).  From these
ratios and from the non-detection of the $\mathrm{H}_2$ lines in the H
spectrum we can assume to have a pure thermally excited spectrum. Two
mechanisms can concour: shocks by either supernovae or nuclear jets and
X-ray heating. The high ratio (1,0)S(3)/(1,0)S(1) would favor the
second possibility at least according to the results from Mouri
(1994). The same result would be inferred from the weakness of the
(2-1)S(3) line according to Draine \& Woods (1990).  In Fig.\ref{exh2}
we plot the excitation diagram of $\mathrm{H}_2$ as derived from our
observations and built following Lester et al. (1988). From the diagram
we can infer an excitation temperature of about 2000 K.  The high
excitation temperature allows us to exclude thermal excitation in
photodissociation regions and favor the thermal process either by SN
driven shocks or X-ray heating.  The (1,0)S(1)/Br$\gamma$ ratio of 5.7
is unusually high even for a Seyfert galaxy (Moorwood \& Oliva 1988).
This makes IRAS~19254-7245 somehow similar to NGC~6240, however, unlike NGC~6240, the
$\mathrm{H}_2$ emission is concentrated on the nucleus of the southern
component and we detect no emission either from the northern nucleus
or from the internuclear region. Though we cannot rule out that the
interaction generates large shocks at least in the south galaxy, there
is no direct evidence that could support the presence of such
phenomena.  Also the (1,0)S(1) luminosity of about $1.8~10^7~L_\odot$
makes of IRAS~19245-7245 a scaled down version of NGC~6240
($10^8~L_\odot$).

Very similarly to $\mathrm{H}_2$, the ratio [FeII]/Br$\gamma$ = 5.5 is
unusually high and again one of the highest after NGC~6240 in the list of
galaxies observed by Moorwood \& Oliva (1988); a similarly high ratio
is reported by Vanzi et al. (1997) for the Seyfert 2 galaxy
Arp~182. Such a high ratio is a clear evidence for AGN in fact values
around unit are expected in starbursting and HII galaxies while values
above unit are typical of AGNs. We can use the ratio [OI]/$H\alpha$=0.17 to locate the 
galaxy in a diagnostic diagram as the one of Alonso-Herrero et al. (1997).
In such a plot IRAS~19254-7245 clearly seats in the Seyfert galaxies region. 
These high ratios are possibly due to the concurrence of
excitation mechanisms such as shocks from SNe, AGN driven shocks and
X-ray heating.

\subsection{Molecular hydrogen mass}
The conversion of integrated line intensity of 
CO to surface mass density of molecular hydrogen involves making
several assumptions (see Morris \& Rickard 1982; Arimoto et al. 1996; 
Bryant \& Scoville 1996; Sakamoto 1996; Verter \& Hodge 1995;
Hollenbach \& Tielens 1997).

The conversion factor is
currently the subject of much discussion, so for consistency many
authors assume the commonly applied ``standard Galactic'' value of
$2.3 \times 10^{20} \mathrm{cm}^{-2} (\mathrm{K~km~s}^{-1})^{-1}$
(Strong et al. 1988), although this value probably overestimates the
true H$_2$ mass in active galaxies. The ``standard'' value can be
compared with independently estimated values from EGRET of $1.56
\times 10^{20}\, \mathrm{cm}^{-2}\, (\mathrm{K~km~s}^{-1})^{-1}$
(Hunter et al. 1997).  Using our SEST flux with the relation for the
conversion of CO line intensity into H$_2$ column density given by
Strong et al. (1988), and an adopted distance of 247~Mpc, we calculate
an {\em indicative } estimate of the total molecular mass of M$_{\mathrm{H}_2} 
= 1.9\times 10^{10}\,\mathrm{M}_{\odot}$ that
is consistent with a similar analysis presented by Mirabel et
al. (1991), but should be regarded as an upper limit. Indeed,
this molecular gas mass is
about an order of magnitude higher than in
typical spiral galaxies (Boselli et al. 1997; Casoli et al. 1998)
and is somewhat larger than largest molecular gas mass ( $1.4\times
10^{10}\,\mathrm{M}_{\odot}$) found for a sample of interacting
galaxies studied by Horellou \& Booth (1997).

\begin{table}
\caption{Molecular hydrogen ratios}
\begin{tabular}{cccc}
\hline
line      &  observed        & fl   & th  \\
\hline
(1,0)S(3) &  1.13 $\pm$ 0.23 & 0.67 & 1.02 \\
(2,1)S(4) &  0.08 $\pm$ 0.08 & 0.12 & 0.02 \\
(1,0)S(2) &  0.35 $\pm$ 0.12 & 0.50 & 0.38 \\
(2,1)S(3) &  0.08 $\pm$ 0.08 & 0.35 & 0.08 \\
(1,0)S(1) &  1.00            & 1.00 & 1.00 \\
\hline
\end{tabular}
\label{h_mod}
\end{table}

\section{Conclusions}

We have observed the Ultra Luminous Infrared Galaxy IRAS~19254-7245,
concentrating our investigation on the southern component with
observations that range from the optical to the millimetre. The
results of our investigation can be briefly summarized as follow:

\begin{enumerate}
   \item Based on our NIR and optical data we estimate 
a visual extinction between 3 and 4 mag. The continuum can be well reproduced assuming this value.
   \item The optical emission lines have broad and complex profiles indicative of gas clouds rapidly moving around the nucleus. The NIR lines are similarly broad though with a much
simpler profile. 
   \item Coronal lines from [FeVII]5721 and [SiVI]1.96 are detected. Their ratio
is consistent, within the error, with the value expected from photoionization. 
   \item NIR lines from $\mathrm{H}_2$ and [FeII] are bright and give unusually high ratios with 
$Br\gamma$ even for the AGN standards. The $\mathrm{H}_2$ is thermally excited and has a very
high luminosity ($1.8~10^7~L_{\odot}$).
   \item From the CO luminosity we derive a $\mathrm{H}_2$ mass of $1.9~10^{10}~M_{\odot}$.
\end{enumerate}

\begin{acknowledgements}
We are grateful to Pierre-Alain Duc for making his NIR images of 
IRAS19254-7245 available to us and to Chad Engelbracht for sending
to us his stellar template spectra. We also thanks Leonardo Testi for
useful discussions and for his careful reading of the first draft of this paper.
RM and ELF acknowledge the support of ESO under the visiting scientist and
studentship programs respectively. We finally thanks the anonymous referee
for very useful comments that improved a lot the present paper.
\end{acknowledgements}

\end{document}